\documentstyle[12pt,epsf]{article}

\newcommand{\beq}{\begin{equation}}
\newcommand{\eeq}{\end{equation}}
\newcommand{\bea}{\begin{eqnarray}}
\newcommand{\eea}{\end{eqnarray}}
\newcommand{\rar}{\rightarrow}
\newcommand{\lan}{\langle}
\newcommand{\ran}{\rangle}

\begin{document}
\font\fortssbx=cmssbx10 scaled \magstep2
\hbox to \hsize{
\includegraphics{uwlogo.ps}
\hskip.5in \raise.1in\hbox{\fortssbx University of Wisconsin - Madison}
\hfill$\vcenter{\hbox{\bf MADPH-95-869}
                      \hbox{February 1995}}$ }
\vskip 2cm
\begin{center}
\Large
{\bf A one parameter representation for the Isgur-Wise function } \\
\vskip 0.5cm
\large
M. G. Olsson and  Sini\v{s}a Veseli \\
\vskip 0.1cm
{\small \em Department of Physics, University of Wisconsin, Madison,
	\rm WI 53706}
\vspace*{+0.2cm}
\end{center}
\thispagestyle{empty}
\vskip 0.7cm

\begin{abstract}
We use a $1S$ lattice QCD heavy-light wavefunction
 to generate a single parameter,
model independent description of the
Isgur-Wise function. Using recent data we find the zero-recoil
slope to be
 $\xi'(1)= -1.16\pm 0.17$, while the second derivative turns out to be
$\xi''(1)= 2.64\pm 0.74$.
\end{abstract}

\newpage
\section{Introduction}
The recent development of the Heavy Quark Effective Theory
 \cite{bib:isgur} yields an  expression for the
$\bar{B}\rar D^{(*)}l\bar{\nu}_{l}$ decay rate in terms of a
single unknown form factor, the Isgur-Wise function (IW). This function
is absolutely normalized at zero recoil point up to corrections of order
$\frac{1}{m_{Q}^{2}}$ \cite{bib:luke}. It is currently
believed that these corrections
can be calculated with less than $5\%$ uncertainty \cite{bib:falk}, which
would allow a precise determination of the CKM matrix element
$|V_{cb}|$ from  the study of
$\bar{B}\rar D^{(*)}l\bar{\nu}_{l}$ decay as a function of the
$D^{(*)}$ meson recoil.

In this note we take a somewhat different approach.
We use an established
LQCD result \cite{bib:duncan,bib:duncan2} for the heavy-light wavefunction
to express the IW function in terms of a single parameter, and then
constrain that parameter using
recent improved experimental results
by the
CLEO II group \cite{bib:cleo} in exclusive
semileptonic $\bar{B}\rar D^{(*)}l\bar{\nu}_{l}$ decay
 \cite{bib:argus}.

\section{Numerical procedure and results}

If one knows the heavy-light meson wavefunction in its rest frame,
and the energy of
light degrees of freedom $E_{q}$, then the IW function for a
given $\omega = v\cdot v'$ (where $v$ and $v'$ are 4-velocities
of two hadrons), can be computed
\cite{bib:sadzikowski} from
\beq
\xi(\omega)
=
\frac{2}{\omega + 1}
\lan\
j_{0}(2E_{q}\sqrt{\frac{\omega-1}{\omega+1}}r)
\ran\ , \label{eq:iwf}
\eeq
where
\beq
\lan A \ran =
\int_{0}^{\infty}dr\  r^2 R(r)A(r)R(r)\ .
\label{eq:norm}
\eeq
{}From the above  the first and   second derivatives
 at the zero recoil
point are \cite{bib:sadzikowski, bib:iwf},
\bea
\xi'(1)& =& -(\frac{1}{2}+\frac{1}{3}E_{q}^{2}
       \lan r^{2}\ran) \ ,\label{eq:first}\\
\xi''(1) &=& \frac{1}{2}+\frac{2}{3}E_{q}^{2}\lan r^{2}\ran
+\frac{1}{15} E_{q}^{4}\lan r^{4}\ran \ . \label{eq:second}
\eea
Higher derivatives could similarly be computed if desired.

The heavy-light wavefunction has been recently computed by a lattice
simulation \cite{bib:duncan, bib:duncan2}. In order to use the
above formulas and the LQCD ``data'', we need an analytic expression
for the wavefunction.  Instead of trying to find some specific
analytic form that would describe the behavior of the lattice data
close to the origin and at large $r$, we  expand the
lattice wavefunction in terms of a complete set of basis
functions. We then truncate the expansion to the first $N$
basis states, hoping that we are be able to find a good
description with a small number of basis states. In other words,
\beq
R_{1S}(r)\simeq \sum_{i=0}^{N-1} c_{i}R_{i}\ .\label{eq:exp}
\eeq
The quasi-Coulombic basis set \cite{bib:weniger} which we have chosen is
particularly well suited for relativistic hadron
systems, and it is given (for s-waves) by
\beq
R_{i}(r)=\sqrt{\frac{8\beta^{3} }{(i+2)(i+1)}}e^{-\beta r}
L_{i}^{2}(2\beta r)\ ,\label{eq:bas}
\eeq
where $L_{i}^{2}$ are  associated Laguerre polynomials and
$\beta$ is a scale parameter. Substituting (\ref{eq:exp}) and
(\ref{eq:bas}) in the expression (\ref{eq:iwf}) yields
\beq
\xi(\omega) =\frac{2}{\omega + 1}
\sum_{i=0}^{N-1}\sum_{j=0}^{N-1}c_{i}c_{j}
\frac{(i+2)!(j+2)!}{\sqrt{(i+2)(i+1)(j+2)(j+1)}}I_{ij} \ ,
\label{eq:iwffin}
\eeq
where
\beq
I_{ij}= \sum_{m=0}^{i}\sum_{n=0}^{j}
\frac{(-1)^{m+n}(m+n+1)!\sin{[(m+n+2)\arctan{(a)}}]}
{m!n!(i-m)!(m+2)!(j-n)!(n+2)!a(1+a)^{\frac{m+n+2}{2}}}\ ,
\eeq
and
\beq
a = \frac{E_{q}}{\beta}\sqrt{\frac{\omega -1}{\omega + 1}}\ .
\eeq
In deriving this expression we have
used the Laguerre polynomial representation
\beq
L_{i}^{\alpha}(x) = \sum_{m=0}^{i}\frac{(-1)^{m}}{m!}
\frac{(i+\alpha)!}{(i-m)!(m+\alpha)!}x^{m}\ .
\eeq

Our radial basis states depend on the meson size parameter $\beta$,
which we estimate from the exponential falloff of the lattice data
($\sim e^{-\beta r}$) to be somewhere between
$0.35$ and $0.45 \ GeV$. Once $\beta$ has been fixed, we vary
the coefficients $c_{i}$ from (\ref{eq:exp}) in order to best
fit the lattice data. Fortunately, for any $\beta$ within the
expected range we are able to find an excellent
approximation to the lattice wavefunction with only 3 basis states, with
$\chi^{2}$ of about $1.1$ per degree of freedom (we have assumed
an uncertainty of 10\% in the wavefunction values \cite{bib:thacker}).
Since all fits were
essentially equivalent, and the wavefunctions were nearly identical up
to $r=20\ GeV^{-1}$, we have chosen
the intermediate value  $\beta=0.40\ GeV$. We emphasize though
that none of the
details of the basis function representation are important to our
final result.
 In Figure \ref{fig:wf} we show our radial wavefunction
for the $1S$ state, together with LQCD data points, for $\beta = 0.40\ GeV$.
As one can see from the figure the agreement is excellent, even
though there is some ambiguity in the data,
especially at large $r$, where the effects
of the small lattice size become large.
The three basis state coefficients ($\beta = 0.40\ GeV$) are found to be
\bea
c_{0} &=& 0.9985\ ,\nonumber\\
c_{1} &=& -0.0221\ ,\label{eq:coef}\\
c_{2} &=& 0.0500\ .\nonumber
\eea

Unfortunately, the energy eigenvalue associated with the lattice
wavefunction is not easily interpreted.
The value of $E_{q}$
depends sensitively upon the lattice spacing \cite{bib:duncan2, bib:thacker}.
Therefore, in the calculation of the IW function we treat
$E_{q}$ as a parameter.

{}From (\ref{eq:iwffin}) and (\ref{eq:coef}) we can now compute $\xi(\omega)$
in terms of $E_{q}$. In Fig. \ref{fig:iwf} we show the
IW function for the range of
\beq
E_{q}=0.306\pm 0.040\ GeV\ ,
\label{eq:eq}
\eeq
which corresponds to  a one standard deviation corridor for the
seven CLEO II \cite{bib:cleo} data points (or $\chi^{2}$ of about
0.65 per degree of freedom).
We also show (full line) the IW function corresponding to the best
fit. Finally, we use (\ref{eq:first}) and (\ref{eq:second}) with
the allowed range of $E_{q}$ (\ref{eq:eq})
to evaluate the first and second derivatives at zero recoil point.
The results are
\bea
\xi'(1)& =& -1.16\pm 0.17\ ,\\
\xi''(1)& = &2.64\pm 0.74\ .
\eea
The best fit corresponding to $E_{q}=0.307\ GeV$, $\xi'(1)=-1.15$,
and $\xi''(1)=2.56$, has
 $\chi^{2}$ of 0.38 per degree of freedom.

\section{Conclusion}

Since the IW function is non-perturbative any parametrization
necessarily must contain some physical input. We have used a reliable lattice
result for the heavy-light meson wavefunction to compute
the IW function in terms of a single parameter $E_{q}$. By comparing
with experimental data we determine the allowed range of this parameter.
Among the direct uses of this parametrization is a more
believable extraction of the IW slope at zero recoil point and
for the first time a reliable estimate of the second derivative.
Previous slope estimates have assumed $\xi(\omega)$ can accurately
approximated by $\xi(\omega)\simeq 1+\xi'(1)(\omega -1)$, but this
inevitably leads to an overestimated (not sufficiently
negative) value for the slope \cite{bib:cleo}.

\begin{center}
ACKNOWLEDGMENTS
\end{center}
We would like to thank
T. Duncan for providing us with the lattice data and H. Thacker for several
useful conversations.
This work was supported in part by the U.S. Department of Energy
under Contract No. DE-FG02-95ER40896 and in part by the University
of Wisconsin Research Committee with funds granted by the Wisconsin Alumni
Research Foundation.

\begin{figure}[p]
\begin{center}
FIGURES
\vskip 2mm
\end{center}
\caption{S-wave radial wavefunction obtained from the fits
to the LQCD data. We used $\beta = 0.40\ GeV$ and $N=3$ basis
states. The wave function is normalized as in (\protect\ref{eq:norm}).}
\label{fig:wf}
\end{figure}

\begin{figure}
\caption{Two limiting cases of the IW function calculated from the
$1S$ radial wavefunction shown on Fig. \protect \ref{fig:wf}, corresponding
to $E_{q}=0.346\ GeV$
and $E_{q}=0.266\ GeV$(dashed lines). The best fit to the
data is for $E_{q}=0.307\ GeV$ (full line).
For the sake of clarity, the error bars are  shown only
for the CLEO II data.}
\label{fig:iwf}
\end{figure}

\begin{figure}
\vskip 9cm
\end{figure}

\begin{figure}[p]
\epsfxsize = 5.4in \epsfbox{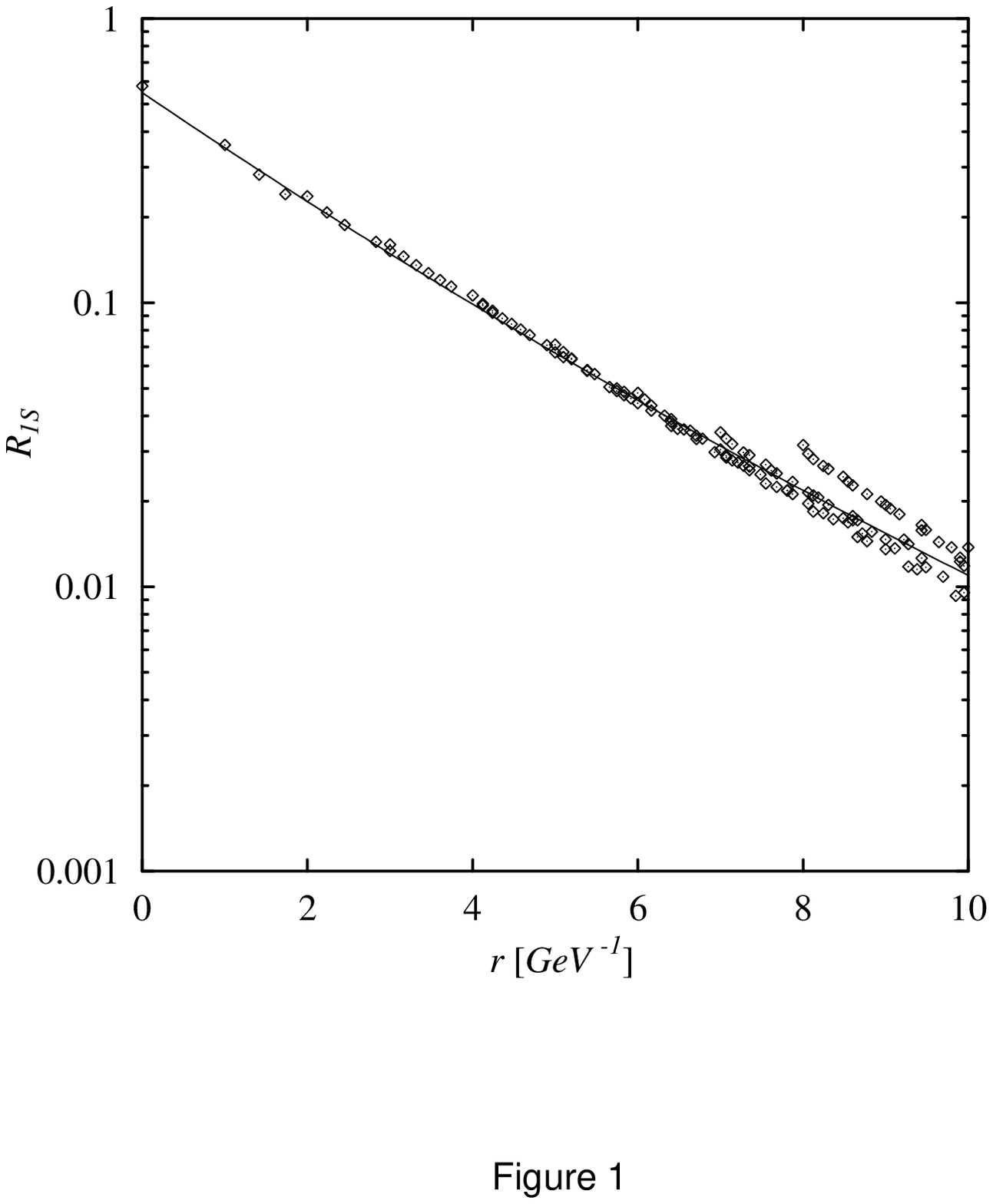}
\end{figure}

\begin{figure}[p]
\epsfxsize = 5.4in \epsfbox{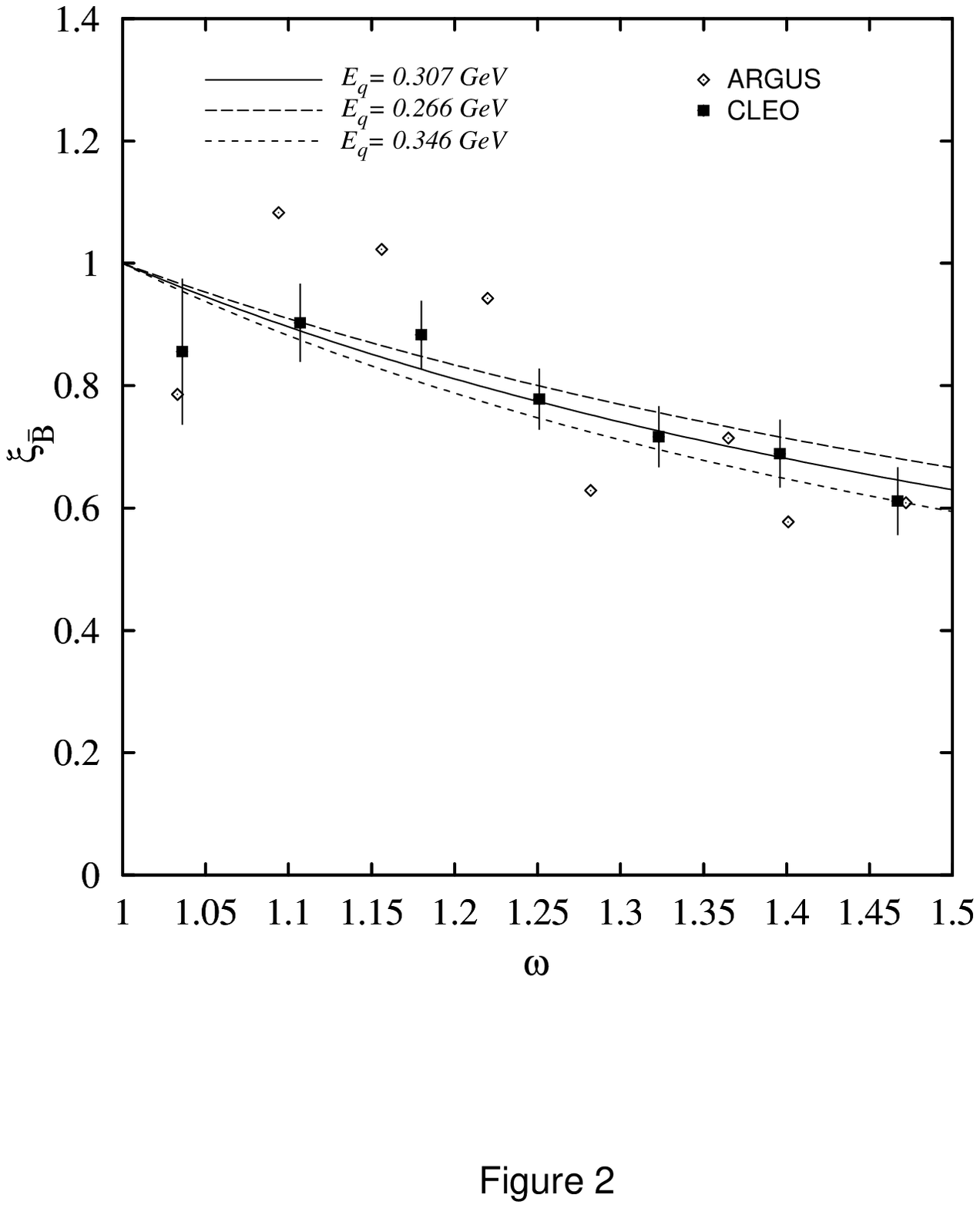}
\end{figure}

\end{document}